\documentclass{article}
\usepackage{icrctc07,graphicx,amssymb}
\title{Transition from galactic to extragalactic cosmic rays}
\shorttitle{Transition from galactic to extragalactic CRs}
\authors{V.~Berezinsky }
\shortauthors{V.~Berezinsky}
\afiliations{INFN, Laboratori Nazionali del Gran Sasso,
I-67010 Assergi (AQ), Italy }%

\email{venya.berezinsky@lngs.infn.it}

\abstract{The transition from galactic to extragalactic cosmic rays is
discussed. One of critical indications for transition is given by 
the Standard Model of Galactic cosmic rays, according to which the
maximum energy of acceleration for iron nuclei is of order of 
$E_{\rm Fe}^{\rm max} \approx 1\times 10^{17}$~eV. At 
$E > E_{\rm Fe}^{\rm max}$ the spectrum is predicted to be very steep 
and thus the Standard Model favours the transition at energy not 
much higher than  $E_{\rm Fe}^{\rm max}$. As observations are concerned
there are two signatures of transition: change of energy spectra and 
elongation rate (depth of shower maximum in the atmosphere $X_{\rm max}$   
as function of energy). Three models of transition are discussed: 
dip-based model, mixed composition model and ankle model. In the
latter model the transition occurs at the observed spectral feature,
ankle, which starts at $E_a \approx 1\times 10^{19}$~eV and is
characterised by change of mass compostion from galactic iron to
extragalactic protons. In the dip model the transition
occures at the second knee observed at energy $(4 -8)\times 10^{17}$~eV
and is characterised by change of mass composition from galactic iron
to extragalactic protons. The mixed composition model describes
transition at $E \sim 3\times 10^{18}$~eV with mass composition
changing from galactic iron to extragactic mixed composition of 
different nuclei. These models are confronted with observational data
on spectra and elongation rates from different experiments, including 
Auger.}
\begin{document}
\maketitle
\section{Introduction}
The Ultra High Energy Cosmic Ray (UHECR) has two most important problems.
One of them is a presence of spectrum features produced by propagation 
of UHECR particles through Cosmic Microwave Radiation (CMB) and the
second is transition from galactic to extragalactic Cosmic Rays (CR). 

In the case of extragalactic  protons two spectral signatures caused 
by interaction with CMB are predicted: Greisen-Zatsepin-Kuzmin (GZK) 
cutoff \cite{GZK}  and pair-production dip \cite{BG88}.

GZK cutoff is most spectacular prediction for UHECR, which status is still
uncertain in present observations, though there are the indications to
its presence. 

The pair-production dip is the spectral feature originated
due to electron-positron pair production by extragalactic
protons  interacting with CMB: $p+\gamma_{\rm CMB}
\rightarrow p+e^++e^-$. Recently this feature has been studied in 
the works \cite{Stanev2000,BGGPL,BGG}. The dip has been observed with 
very good statistical significance $\chi^2$/d.o.f.$\sim 1$ by the Fly's
Eye, Yakutsk, Akeno-AGASA and HiRes detectors, and with much worse
statistical significance by Auger detector. 

The pair-production dip and GZK cutoff are signatures of protons. 
The confirmation of the shape of these features is the evidence for 
proton-dominated composition of primary CRs. For nuclei as primaries 
the shape of the dip and GZK cutoff are strongly modified. 

The different explanation of the dip has been proposed by Hill and 
Schramm \cite{HS85}. They interpreted the dip observed in 1980s 
in terms of two-component model. The low energy component can be
either galactic or produced by Local Supercluster. The similar model   
has been considered in \cite{YT}. 
The Hill-Schramm dip is widely used now for the explanation 
of the observed dip. 

From 1970s in the UHECR spectrum there was observed a flattening,
which is called {\em ankle}. Discovery of this feature at Haverah
Park detector was interpreted as transition from the steep 
galactic component to more flat extragalactic one. The transition at 
ankle has been recently considered in \cite{ankle}. 

In the dip  model the transition is completed at the beginning 
of the dip at $E \approx 1\times 10^{18}$~eV. The ankle in this 
model appears as intrinsic part of the dip. Like in ankle model, 
the transition occurs here 
also as intersection of flat extragalactic component (this flatness 
is especially prominent in case of diffusive propagation) with steep 
galactic spectrum. 

In the dip and ankle models the extragalactic component is assumed to be 
proton dominated, while the galactic component is most probably
composed by iron nuclei. In the {\em intermediate model}, where transition 
occurs in the middle of the dip, the extragalactic CRs are assumed to
have mixed composition \cite{mixed}.

In this paper all three above-mentioned models of transition are
discussed. 
The logic of our discussion is as follows: we approach
first the transition from the high energy end of galactic CRs,  then 
we discuss the properties of UHECR relevant for transition problem and 
finally we describe the transition from properties of these two components.

\section{The end of galactic CRs}

With some disturbing small contradictions one may claim that at
present we have the Standard Model for  Galactic Cosmic Rays. 
It is based on Supernova Remnant (SNR) paradigm and includes four 
basic elements: (i) Supernova Remnants as the sources, (ii) SNR 
shock acceleration, (iii) Rigidity-dependent injection as mechanism 
providing the observed CR mass composition and (iv) Diffusive
propagation of CRs in the galactic magnetic fields.\\*[1mm] 

(i) SNRs are able to provide the observed CR energy production in 
Galaxy, which 
can be found as $Q \approx \omega_{\rm cr} c M_g/x_{\rm cr}$ \cite{book}, where  
$\omega \approx 0.5$~eV/cm$^3$ is the observed CR energy density, 
$c$ is velocity of CR particle, $M_g \approx 5\times 10^{42}$~g is 
the total mass of galactic gas, and $x_{\rm cr} \approx 7$~g/cm$^2$ is the 
grammage traversed by CR before escaping from Galaxy.  Using these 
numbers one obtains $Q \approx 2\times 10^{40}$~erg/s, which is less than 
10\% of energy release in the form of kinetic energy SNR ejecta per 
unit time.\\*[1mm] 

(ii) The great progress has been reached during last decade in the
theory of acceleration. The cosmic ray streaming instability  strongly 
amplifies the magnetic field upstream creating highly turbulent field 
with strength up to 
$\delta B \sim B \sim 10^{-4}$~G \cite{Bell} (for recent works see  
\cite{Blasi}). At each moment of the shock propagation only particles
accelerated to maximum energy $E_{\rm max}$ can escape outside. 
$E_{\rm max}$ reaches the highest value at the beginning of the Sedov 
phase and then diminishes due to shock deceleration. The spectrum of 
escaping particles has a narrow peak at energy 
$E_{\rm max}(t)$ at each moment $t$, but the spectrum integrated
over time has a classical $E^{-2}$ shape with flattening at highest 
energies. This interesting result has been recently obtained by
Ptuskin and Zirakashvili \cite{PZ06}.    

The maximum acceleration energy estimated in the Bohm regime of
diffusion in the  acceleration process is given by 
\begin{equation}
E_{\rm max} = 4\times 10^{15} Z \frac{B}{10^{-4}{\rm G}}
\left ( \frac{W_{51}}{n_g/{\rm cm}^{3}} \right )^{2/5}~~ {\rm eV},
\label{eq:Emax}
\end{equation}
where $B$ is amplified magnetic field, $W_{51}$ is the kinetic energy 
of the shell in units $10^{51}$~erg, $n_g$ is upstream density of the gas   
and $Z$ is charge number of accelerated nuclei. Thus for the protons and
iron nuclei the maximum energies are 
\begin{eqnarray}
E_p^{\rm max} &=& 4\times 10^{15} B_{-4}~{\rm eV},\nonumber\\
E_{\rm Fe}^{\rm max} &=& 1\times 10^{17} B_{-4}~~ {\rm eV} .
\label{eq:pFeEmax}
\end{eqnarray}
$E_p^{\rm max}$ describes well the position of the proton knee and 
$E_{\rm Fe}^{\rm max}$ predicts the position of iron knee. \\*[1mm]

(iii) As the observations show, nuclei are systematically more
abundant in cosmic rays in comparison with interstellar medium in the
solar neighborhood \cite{Ellison,Meyer,BK99}. 
The injection of
particles in the regime of acceleration is responsible for it 
\cite{Ellison,Meyer,BK99}. It can be illustrated by simple 
consideration \cite{Aletal}.   

A particle $i$ from downstream ($i=A,~p$) can cross the shock and
thus to be injected in the regime of acceleration, if its Larmor
radius $r_L(p) \geq d$, where $d$ is the thickness of the shock front. 
Thus we 
readily obtain the relation between nuclei and proton injection momenta
\begin{equation} 
p_{\rm inj}^A = Z e B d /c = Zp_{\rm inj}^p .
\label{eq:p_inj}
\end{equation}
Eq.~(\ref{eq:p_inj}) results in $v_{\rm inj}^A < v_{\rm inj}^p$, which 
provides the higher injection rate of nuclei. 

This conclusion can be reached also in more formal way. Consider flux
of accelerated particles $i$~ $J_i(p)=K_i (p/p_{\rm inj}^i)^{-\gamma_g}$.
Normalizing $J_i(p)$ by condition 
\begin{equation}
\frac{4\pi}{c}\int_{p_{\rm inj}^i}^\infty J_i(p) dp= \eta_i n_i,  
\label{eq:norm}
\end{equation}
where $n_i$ is the density of gas $i$ and $\eta_i$ is a fraction of 
this density injected into acceleration process, we obtain for the ratio 
of fluxes of nuclei and protons in CRs 
\begin{equation}
\frac{J_A(p)}{J_p(p)}=Z^{\gamma_g-1}\frac{\eta_A n_A}{\eta_p n_p}.
\label{eq:fraction}
\end{equation}
Thus, fraction of nuclei is enhanced by factor $Z \eta_A/\eta_p$. 
For numerical calculations of CR nuclei abundances see 
\cite{Ellison,Meyer,BK99}.\\*[1mm]

(iv) CRs propagate in Galaxy diffusively, scattering off small-scale
magnetic turbulence described as superposition of MHD waves with 
different amplitudes and random phases. This process is considered 
(see e.g. \cite{book}) in the resonance approximation, when the
giro-frequency of a particle is equal to a wave frequency in the system
at rest with a motion of a particle along the average magnetic field. 
The magnetic field is separated into average (constant) field 
$\vec{B}_0$ and fluctuating component $\vec{B}$. In \cite{book}
the parallel diffusion coefficient $D_{\parallel}(E)$ is calculated 
assuming $D_{\perp}(E)$ being much smaller than $D_{\parallel}(E)$ 
(see however the numerical simulations \cite{DBS07} which does not 
support this assumption for the highest energies). 

The diffusion coefficient and its energy dependence is primarily 
determined by  spectrum of turbulence $w(k)$ which in most
important cases is given in the power-law form  $w(E) \propto k^{-m}$,
where $k$ is a wave number. Then one has 
\begin{equation}  
 w(k) \propto k^{-m},\;\; D(E) \propto E^{-n},\;\; n=2-m .
\label{eq:D(E)}
\end{equation}
Thus, we obtain for the Kraichnan turbulence spectrum, which Landau and
Lifshitz \cite{LL} consider theoretically preferable for MHD waves, 
$m=3/2$ and $D(E) \propto E^{1/2}$; for the Kolmogorov spectrum 
$m=5/3$ and $D(E) \propto E^{1/3}$ and for diffusion in
shock-dominated turbulence   
$m=2.0$ and $D(E)= const$. In the cases when the turbulent magnetic 
component $\delta B$ is much larger than regular component $B_0$ 
the Bohm diffusion is valid $D(E) \propto E$; this case is in particular
valid for acceleration on the shock fronts. 

Diffusive propagation is the only phenomenon. which imposes
currently the problems for the Standard Model of Galactic CRs. 
The essence of this problem can be easily seen. 

Using the generation spectrum in the Galaxy, as that in acceleration 
$Q_{\rm gen}(E) \propto E^{-2}$, one obtains the diffuse spectrum: 
\begin{equation}
J(E) \propto Q_{\rm gen}(E)/ D(E) \propto E^{-(2+n)} .
\label{eq:diff-problem}
\end{equation}
Then from the observed spectrum $J(E) \propto E^{-2.7}$ one obtains 
$D(E) \propto E^{0.7}$, which in principle  results in too high 
anisotropy  $\delta (E)\propto D(E)$  and 
too low traversed grammage $x_{\rm cr}(E) \propto 1/D(E)$ at high energy E.
There are  suggestions how these problems may be solved:
the problem of small $x_{\rm cr}(E)$ - by spallation inside CR sources
and reacceleration, the problem of anisotropy - by local character of
this phenomenon, and flat spectrum of helium - by acceleration in SNI
remnant enriched by helium (see \cite{BKPZV,BV07} for 
discussion and references).

\begin{figure*}[t]
\begin{center}
\includegraphics [width=0.515\textwidth]{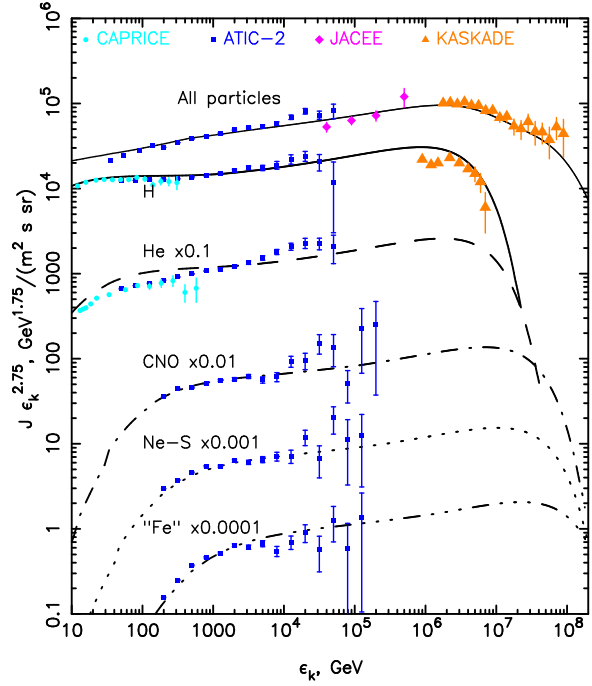}
\end{center}
\caption{Fluxes and spectra calculated within Standard Model in \cite{BV07} 
for all particles, protons and nuclei are shown as
function of kinetic energy $\epsilon_k$. They are compared with data of
CAPRICE, ATIC-2, JACEE and KASCADE. The position of the knees for all
nuclei are given by $\epsilon_{\rm kn} \approx 3Z\times 10^{15}$~eV.
The end of Galactic spectrum is given by iron knee 
$\epsilon_k \approx 8\times 10^{16}$~eV. At higher energies galactic
spectrum becomes very steep.}  
\label{fig:b-volk}
\end{figure*}

The proton and nuclei spectra calculated recently by Berezhko and
V\"olk \cite{BV07}  within Standard Model are shown in Fig.~\ref{fig:b-volk}
in comparison with observational data. The agreement with observations
is quite
good at low energies, and the knee is confirmed at $E_{\rm kn} \approx 
3\times 10^{15}$~eV in proton and all-particle spectra. The iron knee 
located at $E_{\rm Fe} \approx 8\times 10^{16}$~eV is most important
prediction of Standard Model. The spectra beyond the knees are
predicted to be very steep. 
\section{Pair-production dip and GZK cutoff}
\begin{figure}[hb]
\begin{center}
\includegraphics [width=0.40\textwidth]{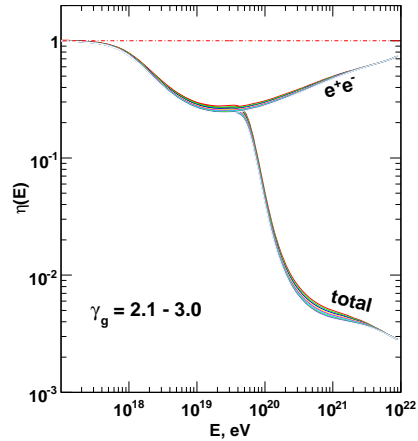}
\end{center}
\vspace{-14mm}
\caption{Modification factor} 
\label{fig:mfactor}
\end{figure}

Being a quite faint feature,
the $e^+e^-$-production dip is not seen well in the
naturally presented spectrum $\log J(E)$ vs.\ $\log E$. The dip is
more pronounced when analyzed in terms of the 
{\em modification factor} \cite{BG88,Stanev2000}. 
It is defined as a ratio of the spectrum $J_p(E)$ calculated with all
energy losses taken into account , and unmodified spectrum 
$J_p^{\rm unm}(E)$, where only adiabatic energy losses are included.
\begin{equation}
\eta(E)=J_p(E)/J_p^{\rm unm}(E)
\label{eq:eta}
\end{equation}
The modification factor is presented in  Fig.~\ref{fig:mfactor}. If one
includes 
only adiabatic energy losses, $\eta(E)=1$ according to
definition (dash-dot line). If $e^+e^-$-production energy losses are
\vspace{1mm}
additionally included one obtains dip, shown in  Fig.~\ref{fig:mfactor} by 
curve ``$e^+e^-$''. If to include the pion production, the GZK feature 
appears (curve ``total''). 
\begin{figure*}[ht]
\begin{center}
   \begin{minipage}[ht]{54 mm}
     \centering
     \includegraphics[width=53 mm]{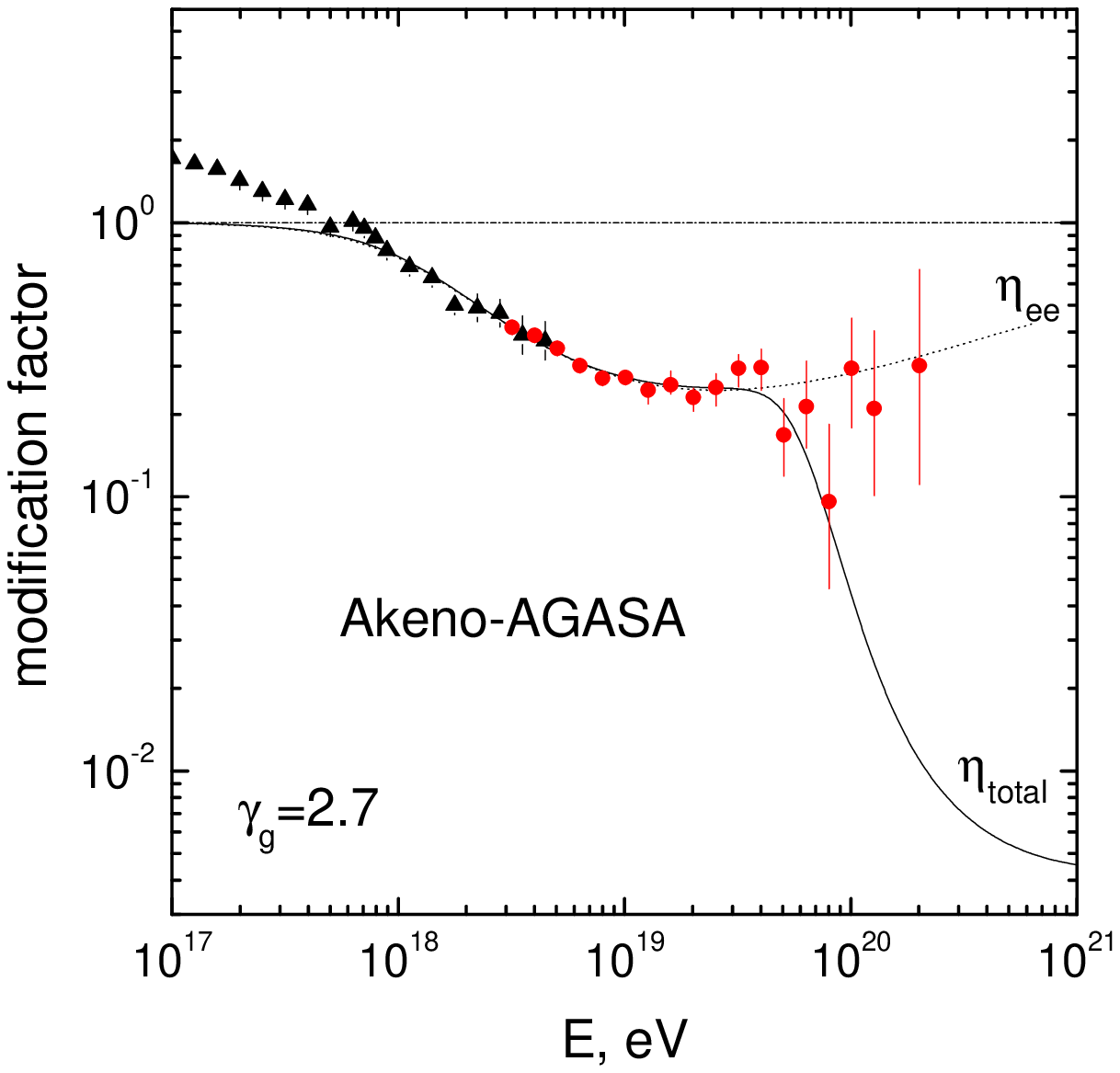}
   \end{minipage}
   \hspace{1mm}
   \vspace{-1mm}
 \begin{minipage}[h]{54 mm}
    \centering
    \includegraphics[width=53 mm]{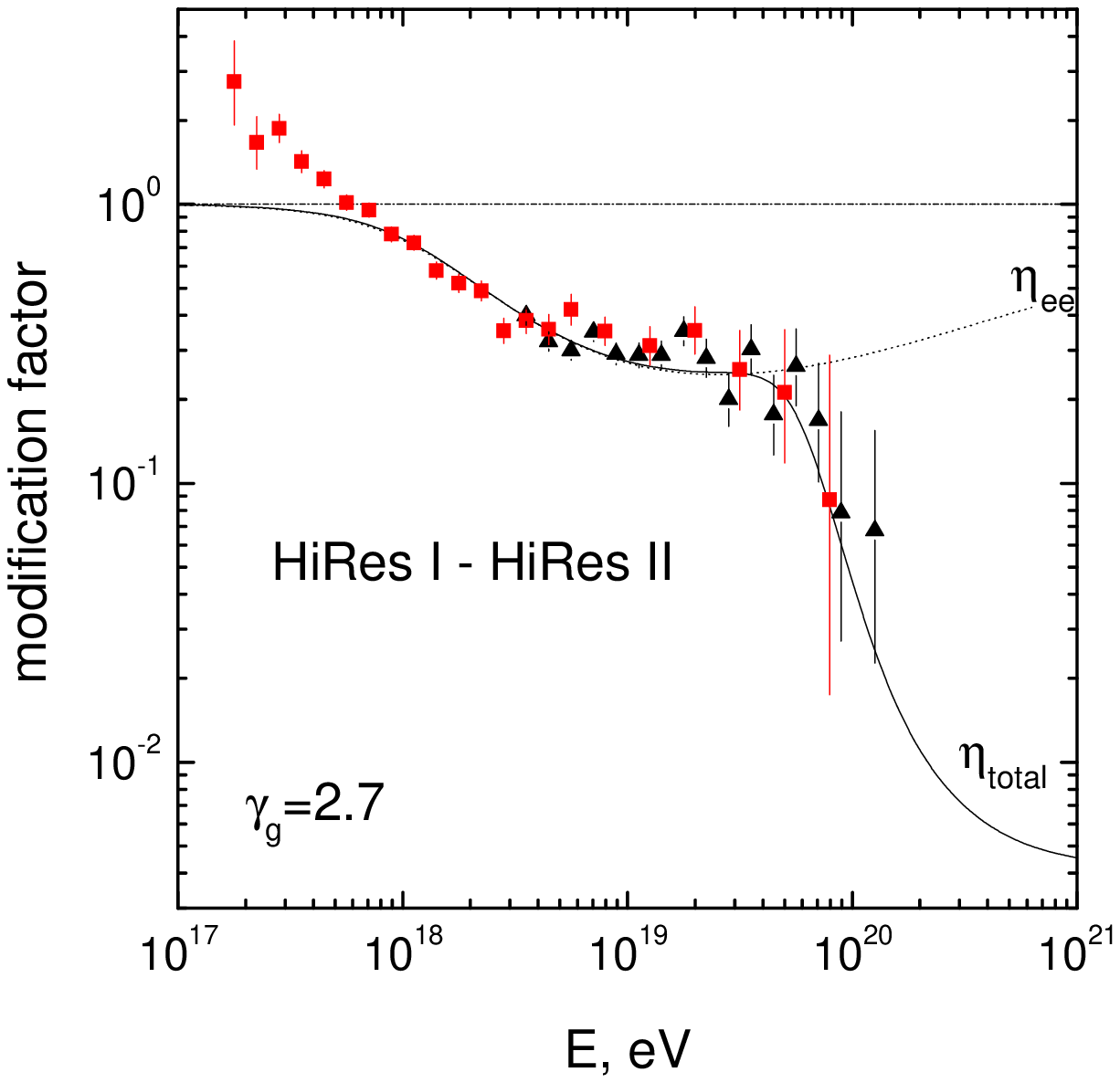}
 \end{minipage}
\medskip
   \begin{minipage}[ht]{54 mm}
     \centering
     \includegraphics[width=53 mm]{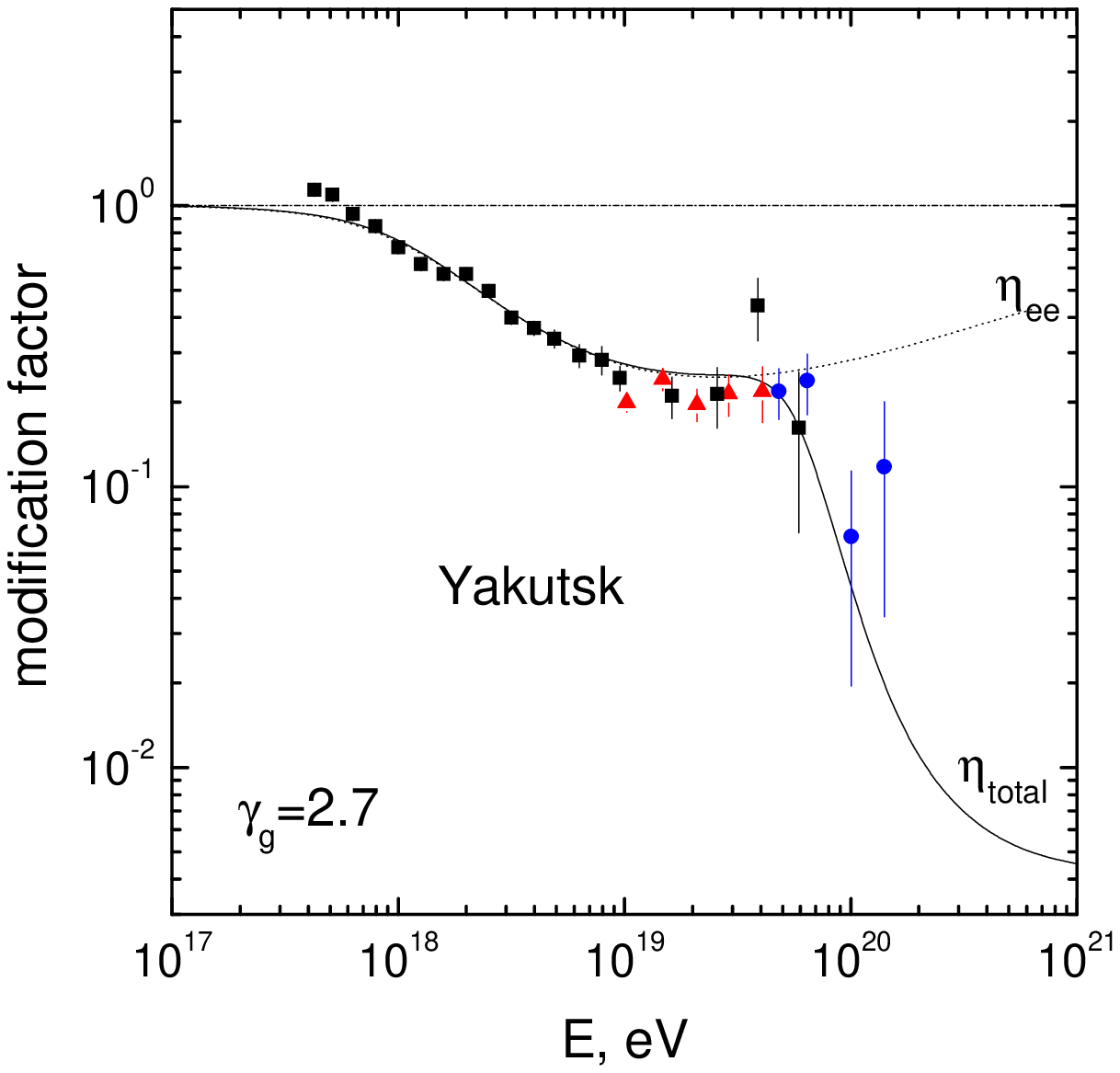}
   \end{minipage}
   \hspace{1mm}
   \vspace{-1mm}
 \begin{minipage}[h]{54 mm}
    \centering
    \includegraphics[width=53 mm]{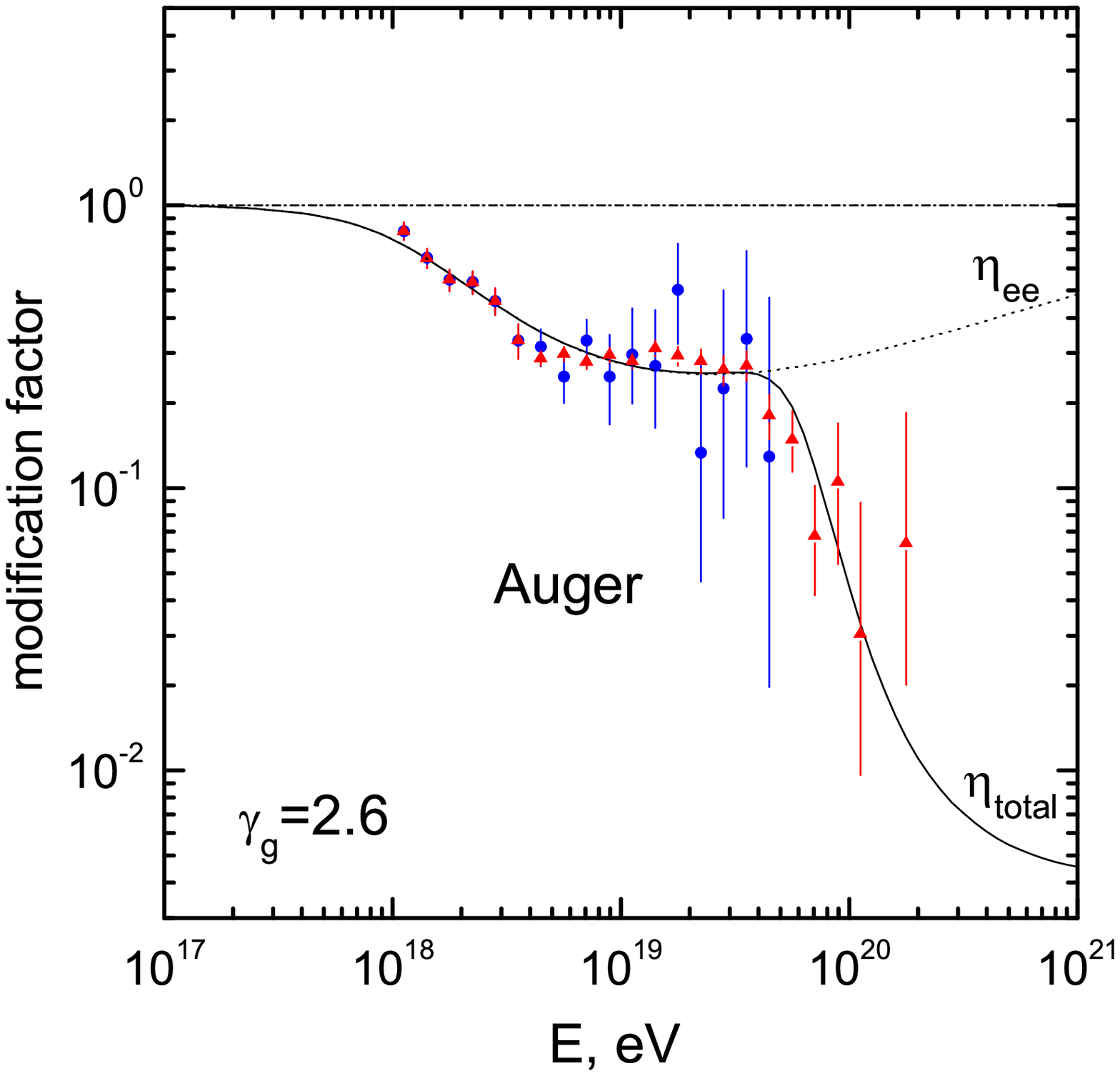}
 \end{minipage}
\end{center}
\vspace{-4 mm}%
\caption{ The predicted pair-production dip in comparison with
  Akeno-AGASA, HiRes, Yakutsk and Auger data \cite{data}.
  The first three experiments confirm dip with good $\chi^2/{\rm
  d.o.f.} \approx 1.0 -1.2$, while the  Auger data are characterized
  by larger $\chi^2/{\rm d.o.f.}$ (see the text).} 
\label{fig:dips}
\end{figure*} %
The observable part of the dip extends from 
beginning of GZK cutoff at $E \approx 4\times 10^{19}$~eV down to 
$E \approx 1\times 10^{18}$~eV, where $\eta \approx 1$. 
It has two flattenings: one
at energy  $E_a \approx 1\times 10^{19}$~eV and the other at  
$E_b \approx 1\times 10^{18}$~eV. The former automatically produces ankle
(see Fig.~\ref{fig:dips}) and the latter provides the intersection of 
flat extragalactic spectrum at $E \leq 1\times 10^{18}$~eV with more
steep galactic spectrum.

The modification factor is less model dependent physical quantity than 
the spectrum.
In particular it depends weakly on spectral index of generation spectrum
$\gamma_g$: In Fig.~\ref{fig:mfactor} the curves are plotted for 
$2.1 \leq \gamma_g \leq 3.0$ with intervals $\Delta\gamma_g =0.1$. The
remarkable property of visible dip in terms of modification factor is 
its {\em universality}.  Modification factor $\eta(E)$ is given as 
dimensionless numbers
for different energies and the curve remains the same when various
physical phenomena are included in calculations \cite{BGG}: 
discreteness in the
source distribution (the distance between sources may change from 
1~Mpc to 60~Mpc), different modes of propagation (from rectilinear to
diffusive), local overdensity or deficit of the sources, large-scale 
inhomogeneities in distribution of sources, some regimes of
cosmological source evolution (most notably those observed for AGN)
and interaction fluctuations. The only phenomenon which 
modifies dip noticeably is presence of more than 15\% of nuclei in 
primary radiation. Therefore the proton  dip in terms  
of modification factor is the universal spectral feature, determined
mostly by interaction with CMB.

The {\em observed} modification factor is given according to definition 
by the ratio of observed $J_{\rm obs}(E)$ to unmodified 
($J_{\rm unm}(E) \propto E^{-\gamma_g}$) spectrum: 
$\eta_{\rm obs} \propto J_{\rm obs}(E)/E^{-\gamma_g}$,
where $\gamma_g$ is the exponent of the
generation spectrum $Q_{\rm gen}(E_g) \propto E_g^{-\gamma_g}$ in
terms of initial proton energies $E_g$.
\begin{figure*}[t]
\begin{center}
\includegraphics[width=0.8\textwidth]{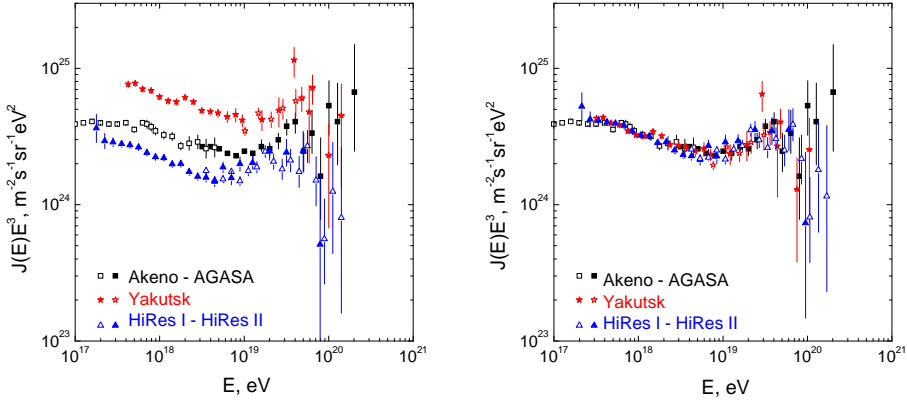}
\end{center}
\caption{The fluxes from  Akeno-AGASA, HiRes and Yakutsk detectors
  before and after calibration.}
  \label{fig:AgHiYa}
\end{figure*}
{\em As  Fig.~\ref{fig:dips} shows the pair production dip and beginning of 
GZK cutoff up to energy $1\times 10^{20}$~eV is reliably confirmed by
all experimental data including AGASA}. As to AGASA excess at 
$E > 1\times 10^{20}$~eV it can be explained by some other  reasons, 
e.g. at some conditions by statistical fluctuations seen in MC of the 
work \cite{DBO2006}.\\*[2mm]
The comparison of the predicted dip with observational data includes only two
free parameters: exponent of the power-law generation
spectrum $\gamma_g$ (the best fit corresponds to
$\gamma_g=2.6 - 2.7$) and normalization constant to fit the
$e^+e^-$-production dip to the measured flux. The number of energy
bins in the different experiments is 20 - 22. The fit is
characterized by $\chi^2/{\rm d.o.f.} = 1.0 - 1.2$ for AGASA,
HiRes and Yakutsk data. For the Auger data $\chi^2$
is good for hybrid data and very bad for surface detector data, mainly 
due to data in two lowest energy bins at 4.3 and 5.5 EeV. In 
Fig.~\ref{fig:dips} the hybrid spectrum shown by circles, and 
combined spectrum (surface detector data combined with fluorescent 
data at low energies) shown by triangles are displayed. If to
introduce the random energy errors $\delta E/E$ inside a bin (see 
section 'Discussion and Conclusions'), which is reasonable for the 
low energy end of the surface detector measurements, $\chi^2$ 
is tremendously improved. The  analysis will be presented 
somewhere else. \\*[2mm]
One can see that at $E < E_b = 1\times 10^{18}$~eV
the experimental modification factor, as
measured by Akeno and HiRes, exceeds the theoretical
modification factor. Since by definition modification factor must
be less than one, this excess signals the appearance of a new
component of cosmic rays at $E < E_b = 1 \times 10^{18}$~eV, and
this component can be nothing but galactic cosmic rays.  
Thus, the transition from extragalactic to galactic cosmic rays,
starts at energy $E_b$.\\*[2mm]
The position and shape of the dip is robustly fixed by interaction
with CMB and can be used for energy calibration of the detectors.

The systematic errors in energy measurements are high, from 15\%
in AGASA to 22\% in Auger. To calibrate each detector we shift the
energies by factor $\lambda$ to reach minimum $\chi^2$ in
comparison with theoretical dip. We obtain these factors as
$\lambda_A=0.9$,~ $\lambda_{Ya}=0.75$~ and $\lambda_{Hi}=1.2$~ for
AGASA, Yakutsk and HiRes detectors, respectively. After energy
calibration the fluxes given by AGASA, HiRes and Yakutsk detectors
agree with each other in a very precise way (see
Fig.~\ref{fig:AgHiYa}). The Auger flux is noticeably below the
flux shown in Fig.~\ref{fig:AgHiYa}.
\begin{figure*}[ht]
\begin{center}
   \begin{minipage}[ht]{56 mm}
     \centering
     \includegraphics[width=55 mm]{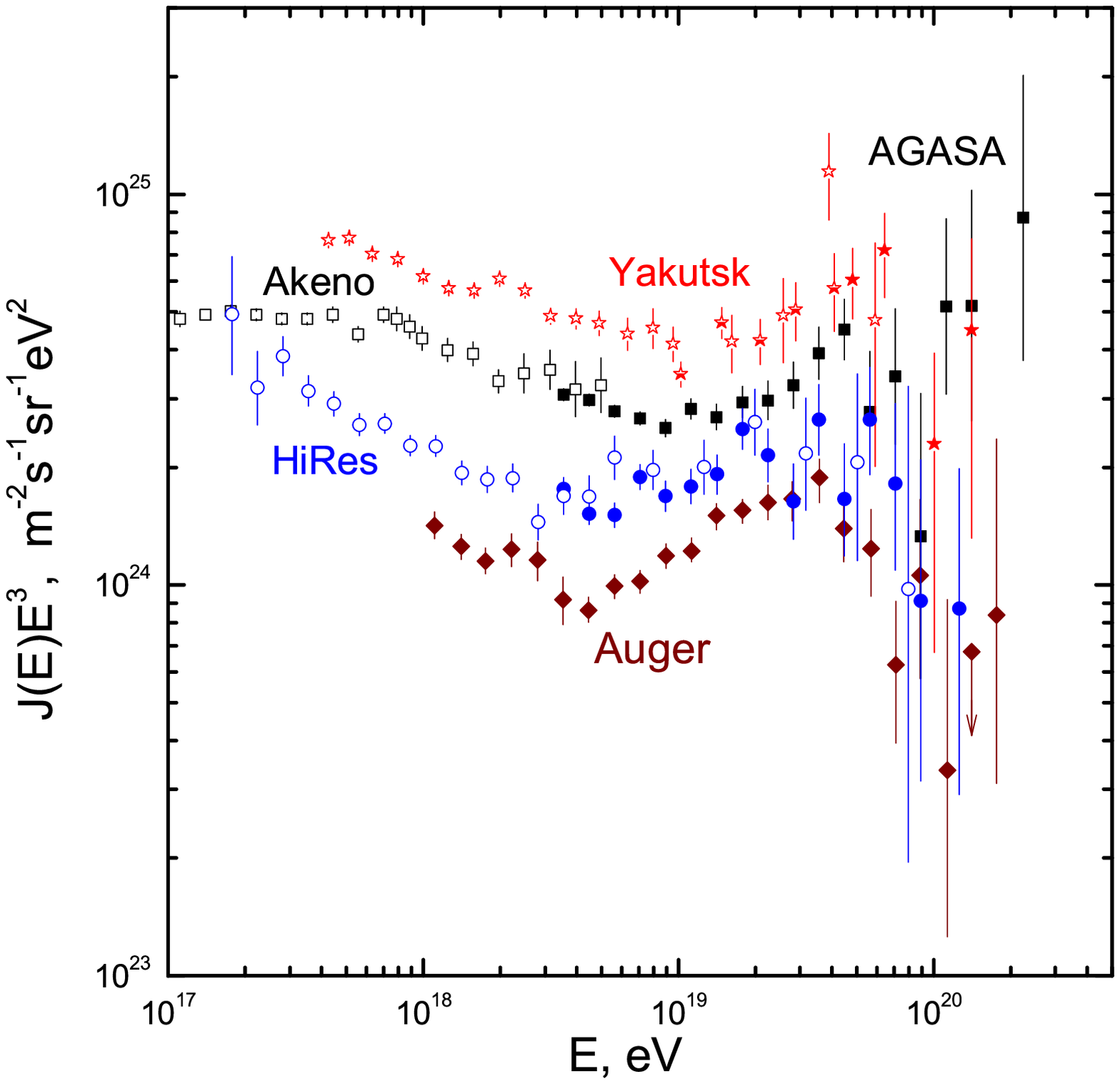}
   \end{minipage}
   \hspace{1mm}
   \vspace{-1mm}
 \begin{minipage}[h]{56 mm}
    \centering
    \includegraphics[width=55 mm]{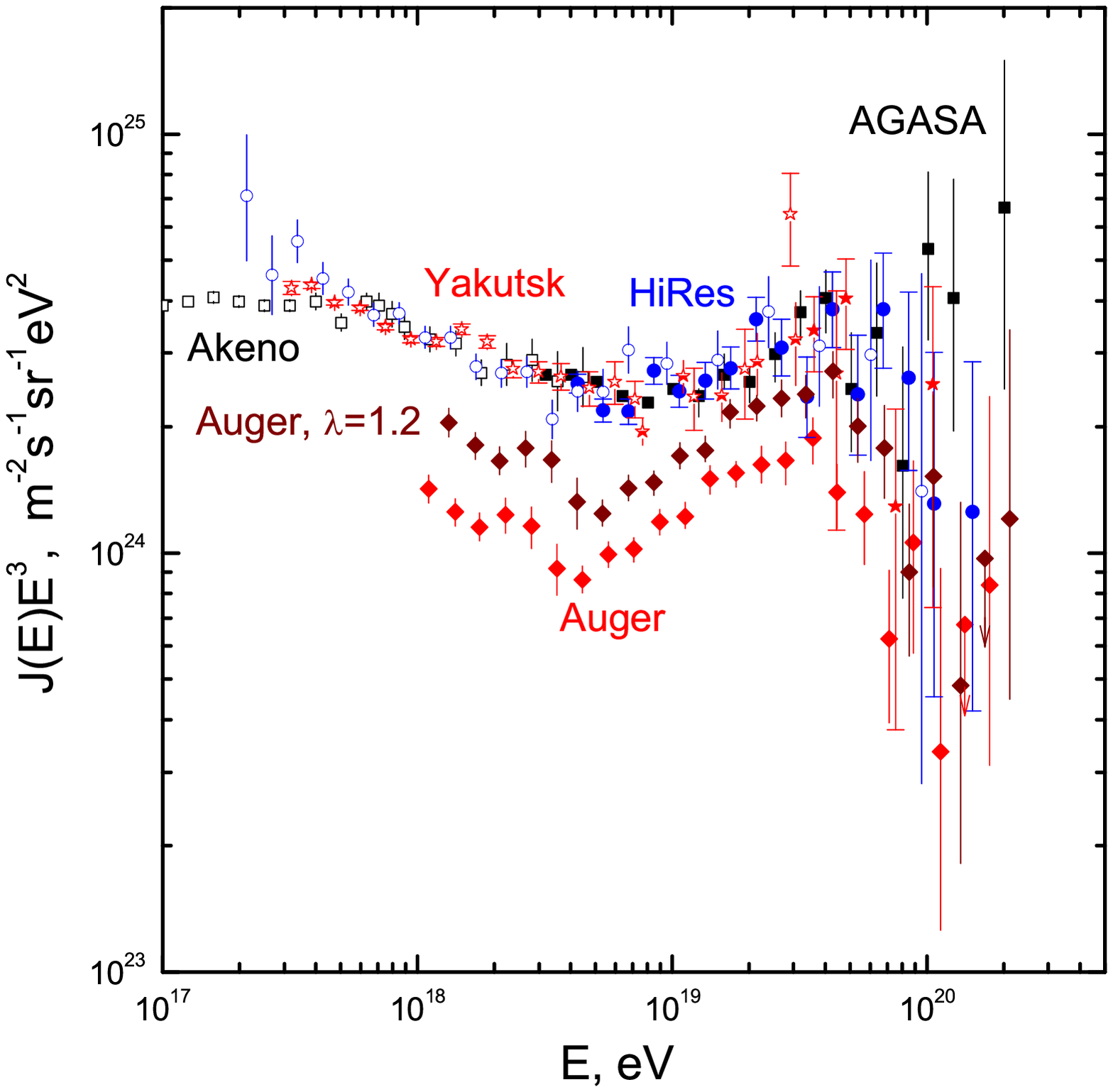}
 \end{minipage}
 \end{center}
\vspace{-4 mm}%
\caption{Comparison of Auger data with AGASA, HiRes and Yakutsk 
(left panel) and comparison of Auger data and the energy shifted 
Auger data ($\lambda=1.2$) with the dip-calibrated  AGASA, HiRes 
and Yakutsk data (right panel).
}
\label{fig:AgHiYaAu}
\end{figure*} %

In Fig.~\ref{fig:AgHiYaAu} we show in the left panel the comparison of 
Auger data with that of AGASA, HiRes and Yakutsk. In the right panel 
we compare the Auger flux with the calibrated data of AGASA, HiRes 
and Yakutsk. We use also the energy-shifted Auger data (curve 
$\lambda=1.2$) with maximum shift allowed by systematic energy 
errors of Auger. One can see that disagreement in fluxes survives.
M. Teshima in his rapporteur talk \cite{Teshima07} noticed that shift with 
$\lambda \approx 1.5$ brings the data of Auger in agreement with the 
calibrated fluxes of AGASA, HiRes and Yakutsk.    
\section{Three models of the transition}
\begin{figure*}[t]
\begin{center}
\includegraphics [width=0.8\textwidth]{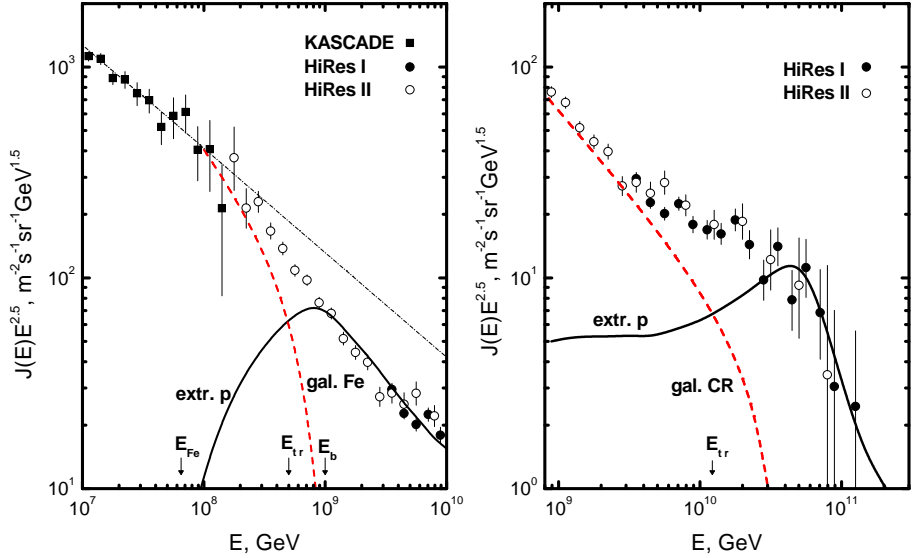}
\end{center}
\caption{Transition in the dip (left panel) and ankle (right panel)
  models. In both cases a solid line gives the calculated spectrum
  of extragalactic protons and a dashed line - spectrum of galactic
  iron. $E_{\rm tr}$ is the energy of intersection of galactic and
  extragalactic spectra and $E_{\rm Fe}$ gives the position of iron
  knee. $E_b=1\times 10^{18}$~eV in the left panel is the energy where
  transition from galactic to extragalactic CRs is completed.
}
  \label{fig:dip-ankle}
\end{figure*}

In this section we describe three models of the transition from 
galactic to extragalactic CRs: {\em ankle}, {\em dip} 
and {\em mixed composition} models. One feature is common for 
all three models: they describe transition as intersection of
steep galactic CR spectrum with more  flat extragalactic spectrum.  
One criterion which all models should respect is agreement with 
the Standard Model of Galactic CRs. The observational data which 
has a power to confirm or reject each model include energy spectrum 
and mass composition.\\*[1.5mm] 
{\em Ankle model}\\*[1mm]
This is a traditional model, based on the interpretation of the ankle as 
spectrum feature of the transition (see \cite{ankle} for the recent
works). In fact this is most natural model,
where transition occurs because extragalactic component is very flat.
This component is assumed to have pure proton composition with flat 
generation spectrum $\propto E^{-2}$  valid for
non-relativistic shock acceleration. Energy losses modify spectrum
insignificantly at $E \lesssim 4\times 10^{19}$~eV.   
The beginning of the ankle $E_a \sim 1\times 10^{19}$~eV corresponds
to equal fluxes of galactic and extragalactic CRs at this energy. 
The transition at the ankle is illustrated by right panel in 
Fig.~\ref{fig:dip-ankle}.  The curve ``extr.p'' presents the calculated 
extragalactic flux of protons and the dash-dot line gives the galactic 
CR spectrum. It is  obtained by subtracting the extragalactic flux from the
total observed flux following the procedure first suggested in \cite{BGH}.
The observed dip in the spectrum is explained not by pair-production
dip, but by Hill-Schramm mechanism \cite{HS85}.  
One must assume that galactic flux is presented by iron nuclei, and
even in this case the ankle model contradicts the Standard Model of 
Galactic CRs, since the half of the observed flux at 
$E \sim 1\times 10^{19}$~eV has the galactic origin. This model needs 
another component of galactic CRs with acceleration to energy 100
times greater than maximum energy in the Standard Model. 

\begin{figure*}[t]
\begin{center}
   \begin{minipage}[ht]{60 mm}
     \centering
     \includegraphics[width=59 mm]{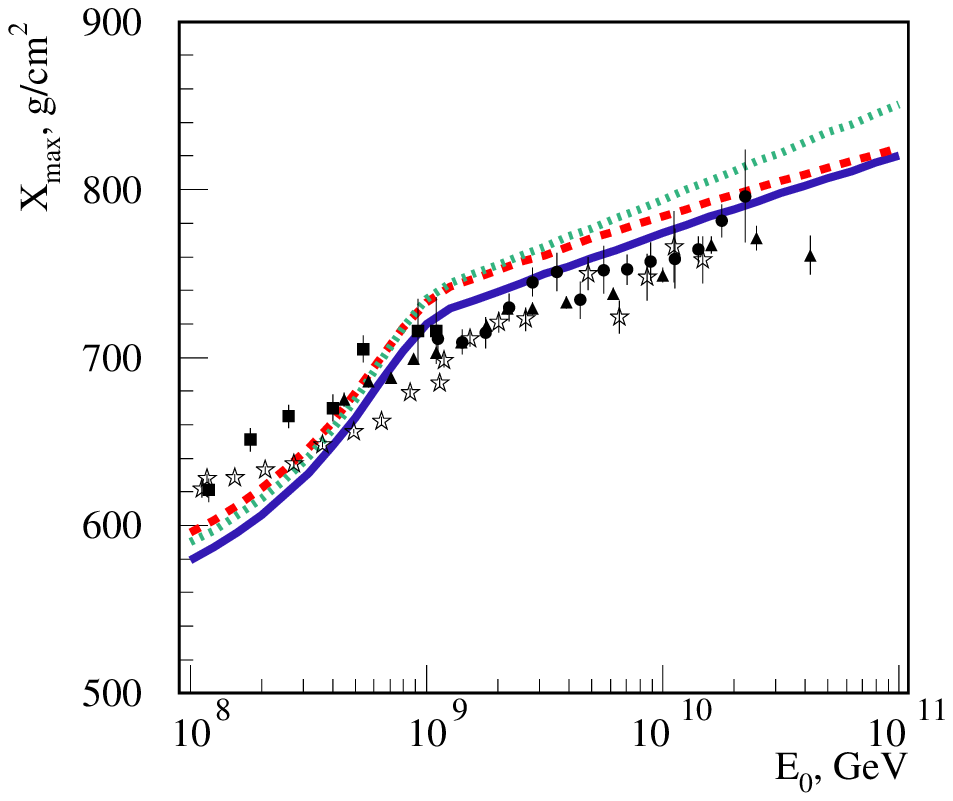}
   \end{minipage}
   \hspace{3mm}
   \vspace{-1mm}
 \begin{minipage}[h]{60 mm}
    \centering
    \includegraphics[width=59 mm]{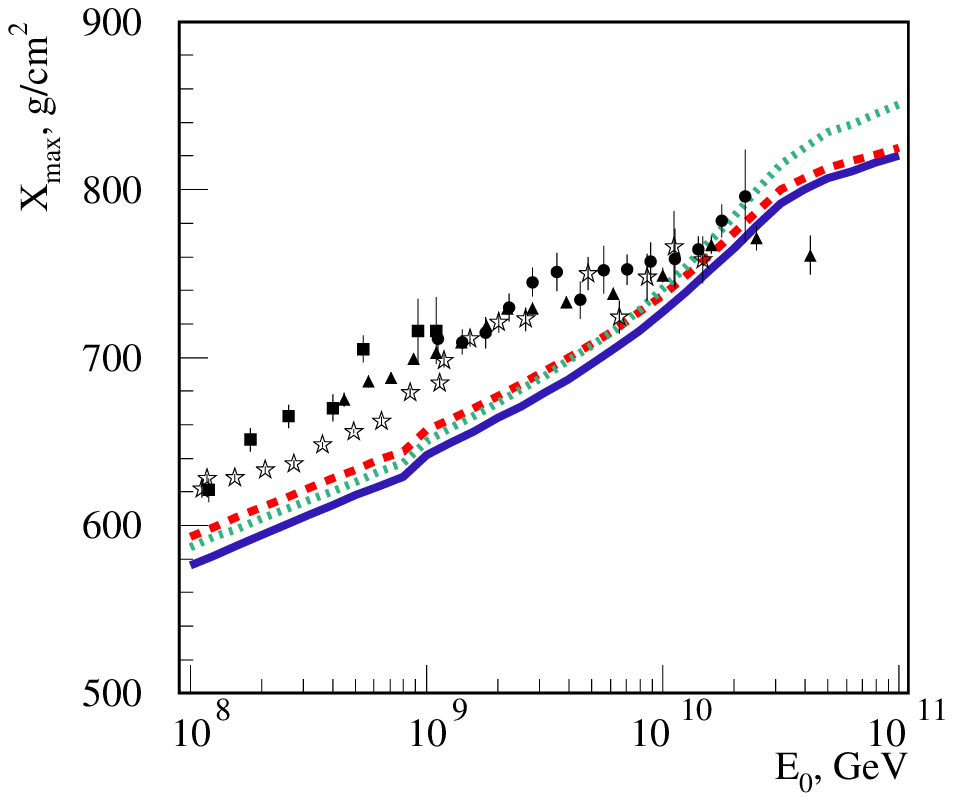}
 \end{minipage}
 \end{center}
\vspace{-4 mm}%
\caption{Elongation rate for the dip model (left panel) and ankle
  model (right panel). The calculated elongation rates are shown by
  the solid lines for QGSJET \cite{QGSJET} model of interaction, by dashed 
  lines for QGSJET-II \cite{QGSJETII}, and by dotted lines for 
  SIBYLL \cite{SIBYLL}. The data points
  are measurements of Fly's Eye (stars), HiRes-Mia (squares), HiRes 
  (circles) and Auger (triangles). 
}
\label{fig:Xmax}
\end{figure*}

Another problem of this model is given by measured {\em elongation rate}
$X_{\rm max}(E)$, where $X_{\rm max}$ is the depth of the atmosphere 
(in g/cm$^2$) where a shower has maximum. In the right panel of
Fig.~\ref{fig:Xmax}  $X_{\rm max}(E)$ calculated for the ankle model
is plotted in comparison with elongation rates measured by different
detectors. One can see that in energy range 
$(1.5 - 5)\times 10^{18}$~eV there is great discrepancy between
elongation rate calculated in all models with measurements of all
detectors \cite{ABBO}.
\newline
{\em Dip model}\\*[0.5mm]
It is based on spectral confirmation of {\em pair-production dip} in 
energy range 
$1\times 10^{18} - 4\times 10^{19}$~eV  and beginning of GZK cutoff 
in energy range $4\times 10^{19} - 1\times 10^{20}$~eV. Since both of
these features are signatures of protons, their observational
confirmation means the indication that mass composition is dominated 
by protons. The shape of the dip allows admixture of nuclei not more
than 10 - 15 \%. The transition from galactic to extragalactic CRs 
is completed at $E_b \approx 1\times 10^{18}$~eV. The appearance of 
galactic CRs at $E \leq E_b$ can be seen from behavior of 
modification factor in AGASA and HiRes experiments below $E_b$ 
(see Fig.~\ref{fig:dips}) and from flattening of calculated spectra 
for both rectilinear and diffusive propagation.  The diffusive
propagation makes flattening of the spectrum below $E_b$ more
pronounced \cite{Le05,AB05,BG07}. One can see this spectrum behavior 
for the case of the Bohm diffusion in Fig.~\ref{fig:dip-ankle} (left 
panel, curve ``extr. p''); the apparent falling-down shape of this curve 
is caused by multiplication of the spectrum by $E^{2.5}$. The
intersection of this curve with the galactic spectrum, shown by dashed 
line, provides the transition from galactic to extragalactic CRs. The
transition occurs at energy $E_{\rm tr} \approx 5\times 10^{17}$~eV. 
The galactic 
component is found by subtracting the calculated extragalactic proton flux
from the observed flux, given by the KASCADE and HiRes data.  
Since the energy of transition $E_{\rm tr}$ is close enough to the position
of iron knee given by Eq.~(\ref{eq:pFeEmax}), the dip model 
fits  perfectly the Standard Model of Galactic CRs.  The galactic
spectrum below the iron knee is presented by iron nuclei, and thus 
transition takes place sharply between iron nuclei and protons.  
The feature in the {\em observed spectrum}, which corresponds to 
the transition in the dip model is the {\em second knee}  
at energy $(4 - 8)\times 10^{17}$~eV as observed in different experiments.
The assumed generation spectrum in the dip model has 
$\gamma_g \approx 2.6 - 2.7$. Being extrapolated to 
$E_{\rm min} \sim 1$~GeV, such spectrum results in too high energy
output of the sources. This problem is naturally solved with an
assumption that the actual source spectrum has a standard shape 
with $\gamma_g = 2.0$ for non-relativistic shocks or 
$\gamma_g = 2.2 - 2.3$ for relativistic ones. However, the natural
distribution of the sources over $E_{\rm max}$ \cite{KS} or 
luminosities \cite{Aletal} results in steepening of 
energy spectrum of generation rate $Q(E_g)$ per {\em unit volume} 
of the universe to larger 
$\gamma_g$, starting from some energy.\\*[2mm] 
 The prediction for elongation rate $X_{\rm max}(E)$ is shown in 
Fig.~\ref{fig:Xmax} (left panel). The characteristic feature of 
the dip model -- sharp transition from galactic iron to 
extragalactic protons -- results in steep increase of 
$X_{\rm max}(E)$ with $E$ below $1\times 10^{9}$~GeV in contrast to 
the ankle model, where the increase of $X_{\rm max}(E)$ is less steep, 
because of very flat proton spectrum at 
$E \lesssim 1\times 10^{18}$~eV. The observational data do not contradict 
the predicted steep increase of elongation rate below $1\times 10^{18}$~eV.

The left panel of Fig.~\ref{fig:Xmax} shows a reasonable agreement of
the dip model with the bulk of experimental points in this figure, 
especially if one takes into account $20 - 25$~g/cm$^2$ of systematic 
error in all experiments. However, the detailed comparison of the dip 
prediction with the data of each experiment shows the different
picture. While elongation rate predicted in the dip model agrees well 
with HiRes and HiRes-Mia data, it does not agree with the Auger data 
especially with two highest energy points.\\*[2mm] 
{\em The mixed composition model}\\*[1mm]
The main concept of the mixed composition model  (see Allard et al from 
\cite{ankle}, \cite{mixed}, \cite{Allard07}, \cite{Globus07}) is 
based on the argument
that any acceleration mechanism operating in the gas involves the
different nuclei in acceleration process and thus the primary flux
must have mixed composition. For injection into process of
acceleration the authors assume A-dependent regime, instead of 
rigidity-dependent one (\ref{eq:p_inj}) in the Standard Model, and obtain
$J(E) \propto A^{\gamma_g-1}E^{-\gamma_g}$ instead of Eq.~(\ref{eq:fraction}) 
valid for the Standard Model. It results in higher abundance of CRs by 
heavy elements in comparison with the Standard Model. In fact, as
discussed in \cite{Aletal}, there are the reasonable regimes of injections
when abundance of heavy nuclei is suppressed. 
\begin{figure*}[t]
\begin{center}
   \begin{minipage}[ht]{63 mm}
     \centering
     \vspace{-2mm}
     \includegraphics[width=69 mm, height=55.5 mm]{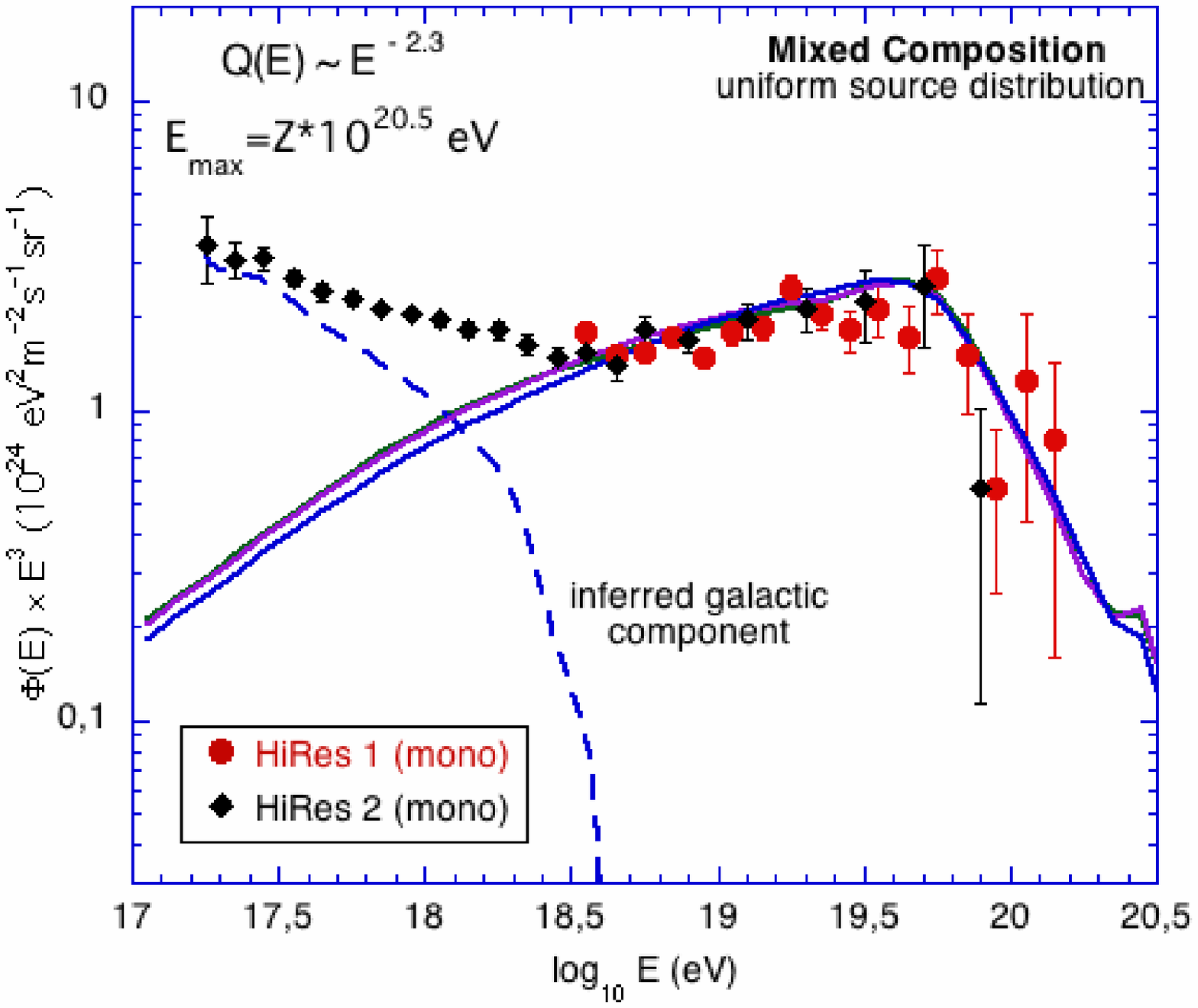}
   \end{minipage}
   \hspace{3mm}
 \begin{minipage}[h]{63 mm}
    \centering
    \includegraphics[width=69 mm, height=54.5 mm]{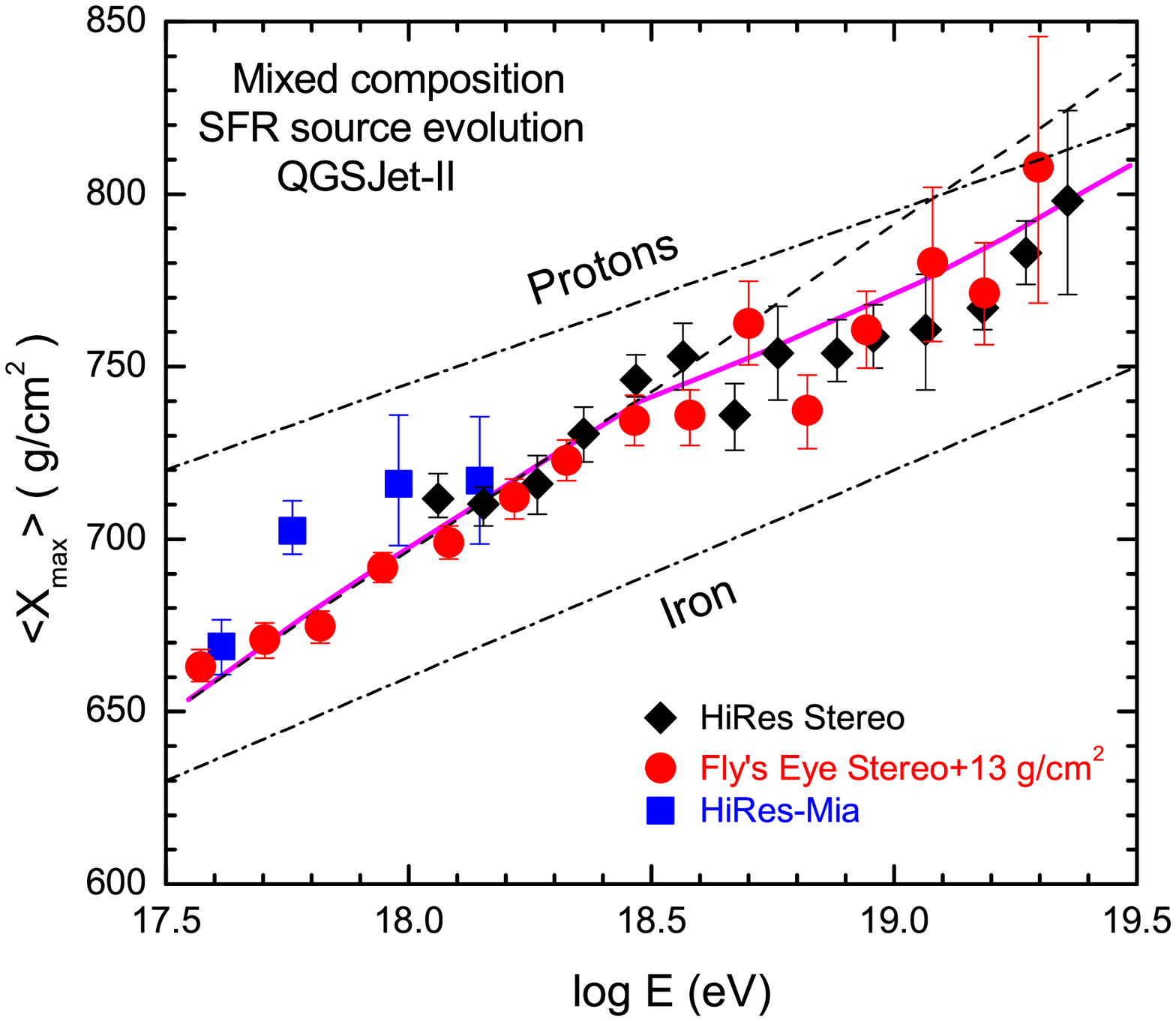}
 \end{minipage}
 \end{center}
\vspace{-4 mm}%
\caption{The spectrum (left panel) and elongation rate (right panel)
for the mixed model \cite{Allard07} with $\gamma_g=2.3$, 
$E_a=3\times 10^{18}$~eV, cosmological source evolution $(1+z)^3$ 
at $z \leq 1.3$ and a set of parameters $x_i$  (see the text).
At $E > 3\times 10^{19}$ eV the spectrum is strongly proton-dominated
and is characterized by GZK cutoff (left panel). The mass composition 
evolves from almost pure iron composition at 
$E \approx 3\times 10^{17}$~eV to the lighter composition due to
enrichment by protons and light nuclei of extragalactic origin. 
At energy $E_a=3\times 10^{18}$~eV the transition to pure
extragalactic component is completed and chemical composition evolution
proceeds further  due to photo-disintegration of the nuclei. At energy 
$E \approx 1.3\times 10^{19}$~eV, seen in the plot, all nuclei are   
disappearing faster than before and composition becomes strongly
proton-dominated at $E \geq 3\times 10^{19}$~eV.
}
\label{fig:mixed}
\end{figure*} %
The UHE extragalactic nuclei propagating through infra-red (IR) and 
CMB radiation are efficiently photo-disintegrated starting with energy 
$E \sim 1\times 10^{19}$~eV, while protons survive and therefore GZK 
feature is present in the mixed-composition model. At energy below 
$1\times 10^{17}$~eV the authors consider the mixed-composition
spectrum \cite{Allard07} which is proportional to injection spectrum 
in the Galaxy:
\begin{equation}
Q_i(E)= x_iA_i^{\gamma_g-1}K E^{-\gamma_g},
\label{eq:mix}
\end{equation}
\vspace{1mm}
where $K$ is a normalization constant, $i$ is a type of nuclei,  
$x_i$ are free parameters, which describe the source chemical composition, 
and $\gamma_g$ is a spectral index, chosen to fit the data, with 
preferable values between 2.1 - 2.3,  motivated by acceleration at 
the relativistic shocks. The cosmological evolution of the sources 
are included in calculations using factor $(1+z)^m$ up to $z_{\rm max}$ 
with different $m$, including $m=0$, and different $z_{\rm max}$.\\*[2.5mm]  
With (\ref{eq:mix}) taken as generation spectrum, the
authors calculate the diffuse spectrum at higher energies 
propagating protons and nuclei  through IR and CMB radiation
from the sources distributed uniformly  in the universe.  Using the 
calculated spectrum they fit
the observed spectrum at energy higher than $E_a=3\times 10^{18}$~eV,
which is thus the energy where the pure extragalactic spectrum starts,
i.e. transition is completed. 

The galactic component is found by subtraction of calculated
extragalactic spectrum from the total observed spectrum. This
procedure, adopted from \cite{BGH}, gives the spectrum below $E_a$ 
as observed and provides the smooth transition to calculated
extragalactic spectrum at $E_a$. Therefore, the part
of the observed dip below $E_a$ is reproduced in this procedure 
phenomenologically in contrast to the pair-production dip, which is
accurately calculated.  

The calculated spectrum and mass composition depends on parameters 
$x_i$ in Eq.~(\ref{eq:mix}), $\gamma_g$,  parameters of cosmological 
evolution $m$ and $z_{\rm max}$, and therefore it is most flexible
model among the three models at the discussion, which is able in
particular to reproduce an arbitrary mass composition with some 
exception. The robust prediction for spectrum and
mass composition is related to energy range $E > 1\times 10^{19}$ eV, where 
the fraction of protons becomes large and steadily increasing, resulting 
thus in the GZK feature and almost pure proton composition, 
in contradiction with recent results of Auger. 

The spectra and mass composition predicted in one of the versions of the
mixed model \cite{Allard07} are displayed in Fig.~\ref{fig:mixed}.
The mass composition is in a good agreement with the selected data of 
Fly's Eye (only stereo), HiRes (only stereo) 
and HiRes-Mia, shown in the figure. The predicted 
elongation rate has two break points, the first at 
$E_a = 3\times 10^{18}$~eV, and the second at 
$E \approx 1\times 10^{19}$~eV. The first one occurs 
when transition to extragalactic CRs is completed and evolution
continues due to photo-disintegration  of nuclei, first iron, then CNO
and finally helium. At energy $E \geq 1\times 10^{19}$~eV (seen in the
figure as $E=1.3\times 10^{19}$~eV)  all nuclei are destroying faster 
and at $E \geq 3\times 10^{19}$~eV the composition becomes strongly 
proton dominated with GZK feature in the energy spectrum.

The first break point agrees with the Auger feature in elongation
rate, but prediction of increasing $X_{\rm max}$ at the second break
point, i.e. at
$E > 1\times 10^{19}$~eV, contradicts to  decreasing of $X_{\rm max}$ 
in the  Auger data. 

\section{Discussion and Conclusions}
\vspace{1.3mm}
The region of transition from galactic to extragalactic CRs at 
energy between $1\times 10^{17} - 1\times 10^{19}$~eV is the key 
energy range for understanding the origin of CRs. At low energy part 
it includes the high energy end of galactic CRs. The information on  
maximum energy of acceleration, chemical composition and propagation 
in Galaxy at these energies will clarify the total picture of origin 
at lower energies. 
The low energy part of UHECRs is important for understanding of 
origin of UHECRs and their propagation in extragalactic magnetic fields.   
The transition from galactic to extragalactic CRs is the central issue
of this energy region. 

There are two detectors which cover partially the above-mentioned region: 
KASCADE-GRANDE \cite{KASCADE-G} and TALE \cite{TALE}. There are also
the proposals to extend the observations of Auger to energy $E \sim
1\times 10^{17}$~eV (see e.g. \cite{LE-Auger}). The Auger detector has
great potential to explore this region, building more dense part
of the detector covered with fluorescent, scintillator and muon detectors. 

The basic information which can be obtained includes precise
measurement  of energy spectra and mass composition (there is little
hope to detect anisotropy in this energy region, though in some models 
the galactic sources can be observed in protons with energy 
$E \lesssim 10^{18}$~eV \cite{BGH}). 

At present we have the sufficiently good data on spectra and mass 
composition at energy range $1\times 10^{18} - 4\times 10^{19}$~eV. 
The spectra are measured with high statistics (especially in case 
of the Auger detector), but problem is the accuracy of energy
determination. From 
quite disappointing Fig.~\ref{fig:AgHiYaAu} (left panel) one concludes that 
scales of energy determination is quite different in all detectors.    
Energy calibration with help of the pair-production dip suggests that 
energy measured by scintillator detectors is
systematically higher than that by the fluorescence detectors and it 
gives a reasonable recipe of increasing energies given by
fluorescent method and decreasing it for the  scintillation
method. In this case the curves 'Yakutsk' and 
'Akeno-AGASA' in Fig.~\ref{fig:AgHiYaAu} go down and 'HiRes' and 'Auger'
- up. For HiRes, AGASA and Yakutsk the method of calibration
with help of dip works successfully (see Fig.~\ref{fig:AgHiYa})
with energy shift within the allowed systematic errors,
but for Auger it requires the shift by factor 2 greater than systematic 
error.  

The pair-production proton dip in terms of modification factor is an
excellent tool to measure {\em spectrum shape} independently of
absolute flux. From Fig.~\ref{fig:dips} one sees the excellent agreement
of the theoretical dip with data of AGASA, HiRes and Yakutsk. By  the 
standards of cosmic-ray physics the agreement with Auger data is also 
good, but $\chi^2$ for comparison with SD data is very large. This is 
a result of very big statistics in the surface detectors at lowest 
energies $E \geq 4.5\times 10^{18}$~eV. In the lowest energy bin 
at $E=4.5\times 10^{18}$~eV there are
4128 events and the error in determination of flux provided mostly by this
statistics is $\delta J/J=0.024$. The theoretical value of modification
factor at this energy is only 14\% higher than experimental value, but
owing to very small $\delta J/J$,the contribution of this bin to $\chi^2$ 
is 99.27 ! Most probably the other sources of errors should be
included in the bins with small $\delta J/J$, and a possible source of 
this error is the energy errors which are changing randomly inside a bin. 
These could be statistical errors and energy-dependent part of
systematic errors. Assuming that number of events are distributed  
in a bin as $N(E)=K E^{-\gamma}$ one obtains 
$\delta J/J = \gamma (\delta E/E)_r$, where $(\delta E/E)_r$ is the
random energy error inside the bin. The estimated value $\delta J/J$
is much larger than what obtained in Auger analysis for all reasonable 
values of $(\delta E/E)_r$ and $\gamma$. More generally, according to 
Markus Roth's remark, $\chi^2$ analysis is not adequate for the cases of 
small  $\delta J/J$ and large $(\delta E/E)$. At this stage of
analysis we do not consider Fig.~\ref{fig:dips} as contradiction with
Auger data.\\*[4mm] 
Coming to the transition from galactic to extragalactic CRs, we 
emphasize that at present there are only two experimental methods to
study it: measuring the spectrum and mass composition. The transition 
will be clearly seen if spectrum of iron nuclei and that of protons are  
measured separately (see Fig.~\ref{fig:dip-ankle}), but even without 
this ideal possibility the total spectrum has signatures of transition 
in the form of the spectral features - {\em second knee} in case of the dip 
model and {\em ankle} in case of the ankle model. The spectrum can be 
measured nowadays with high accuracy and its shape contains the
information about mass composition, which is the other characteristic
of the transition. The pair-production dip with its specific shape is
a signature of proton-dominated composition (nuclei contribution
should be not more than 10 -15 \% \cite{BGGPL}) and its observational
confirmation is an argument not weaker than that due to 
$X_{\rm max}$ measurement (we remind that only two free parameters 
are involved in describing about 20  energy bins in each experiment).
\\*[4.1mm] 
The mass composition gives another way to test the transition. 
The best method at present is given by measuring of elongation 
rate $X_{\rm max}(E)$. Unfortunately this method has many
uncertainties, including those in value of fluorescent yield, absorption of 
UV light in the atmosphere and uncertainties in the models of 
interactions, needed to convert the tested mass composition into  
$X_{\rm max}$. The systematic errors in measuring $X_{\rm max}$ can 
be as large 30~g/cm$^2$ to be compared with difference about 
100~g/cm$^2$  between $X_{\rm max}$  for protons and iron.  The better 
sensitivity for distinguishing
different  nuclei is given by distribution over $X_{\rm max}$ 
\cite{ABBO}.

There are three models of the transition: ankle, dip and
mixed-composition model. They differ  
most notably by the energy of transition (ankle: 
$E \sim 1\times 10^{19}$~eV, dip: $E \approx 1\times 10^{18}$~eV and mixed 
composition model $E \approx 3\times 10^{18}$~eV), and by mass
composition of extragalactic component (protons -  for the ankle model,
proton-dominated -  for the dip model and mixed composition -  for the third
model). 

The {\em ankle model} contradicts the Standard Model of Galactic CRs 
(energy where galactic flux is half of that observed is two orders of
magnitude higher than energy of iron knee) and severely disagrees with 
$X_{\rm max}$ measured in all experiments at $(1.5 - 5)\times 10^{18}$~eV. 

The {\em dip model} is based on well confirmed signature of proton 
interaction with CMB - pair-production dip. The two other models must 
assume that agreement of pair-production dip with data is accidental 
and the observed dip is produced by two components, galactic and 
extragalactic. The dip model assumes the iron-dominated galactic flux 
below $5\times 10^{17}$~eV and proton-dominated extragalactic flux 
above $1\times 10^{18}$~eV. This mass composition is confirmed by 
HiRes and HiRes-Mia data for elongation rate. It does not contradict 
the bulk of all data on $X_{\rm max}$, but contradicts $X_{\rm max}$   
measured by Auger, especially the highest energy points. The 
generation spectrum in this model is $E^{-2}$ or $E^{-2.2}$ as needed  
by shock acceleration with a steepening to $\gamma_g=2.7$ due to
distribution of sources over maximum energy of acceleration of 
source luminosities. The proton-dominated composition can be 
produced in some models of injection to the shock acceleration.   

The {\em mixed composition model} assumes mixed composition 
generation spectrum for extragalactic component with generation 
index 2.1 - 2.3. It has many free parameters, most notably ones
describing the mass composition of the generation spectrum, and 
thus it can in principle explain 
any observed mass composition. However, this model has a
robust prediction at energy $E \gtrsim 3\times 10^{19}$~eV:
proton-dominated composition and the GZK feature. As far as Auger 
elongation rate is concerned, the mixed composition model explains 
well the break 
in elongation rate at $2\times 10^{18}$~eV and contradicts the two
Auger points at $E > 2\times 10^{19}$~eV. The energy where transition 
to extragalactic CRs is completed in most versions of this model 
equals $E \approx 3\times 10^{18}$~eV. Much better quality of data 
on $X_{\rm max}$ is needed to distinguish the dip and mixed-composition
models by $X_{\rm max}$ measurements. Probably it is possible to do
using $X_{\rm max}$ distribution \cite{ABBO}.\\*[1mm]  
We will comment now on agreement of the transition models with the 
measured galactic spectrum. For all three models it is reached by 
the formal subtraction procedure: the galactic spectrum is found as 
difference between measured total spectrum and calculated
extragalactic spectrum. But the galactic spectrum calculated in 
the Standard Model at $E \gtrsim 1\times 10^{17}$~eV is very steep
and, as was demonstrated in \cite{Tanco}, for diffusive model of
propagation  all three models contradict the calculated galactic
spectrum, the dip model to the less extent. Strictly speaking this 
contradiction is produced by exponential cutoff in the acceleration 
spectrum at $E > E_{\rm max}^{\rm acc}$. \\*[2mm]
The most consistent conclusions on nature of observed  UHECRs are obtained   
at present by HiRes detector: it has confirmed the pair-production dip and thus 
proton-dominant composition at $1\times 10^{18} - 4\times 10^{19}$~eV, the 
$X_{\rm max}$ measurements agree with proton-dominant composition 
at $E > 1\times 10^{18}$~eV , and $E_{1/2}$ measurement confirms that 
steepening of the spectrum observed at $E > 4\times 10^{19}$~eV is
really the GZK cutoff. Therefore, according to these data CRs observed
at $E \gtrsim 1\times 10^{18}$~eV are extragalactic protons exhibiting
two signatures of interaction with CMB: pair-production dip and 
GZK feature. 
\section*{Acknowledgments}
I am grateful to Askhat Gazizov for the joint work on analysis of 
Auger data and for numerous discussions. I acknowledge the illuminating
discussions about data errors with Markus Roth, Gianni Navarra and 
Karl-Heinz Kampert. My collaborators      
R. Aloisio, P. Blasi and S. Ostapchenko are thanked  for joint
work on related subject and for many discussions.  
This work is supported in part by \emph{ASI} through grant \emph{WP 1300} 
(theoretical study).

\end{document}